\documentstyle[12pt]{article}
\vspace{1.8cm} \oddsidemargin 0pt \evensidemargin 0pt  \topmargin
-1.5cm \footheight 26pt \footskip 10mm

\begin{document}
\baselineskip 20pt
\begin{center}
\baselineskip=24pt {\Large \bf Non-factorization contributions in
$D\rightarrow \pi K, KK$ decay}

\vspace{1cm} {Xiang-Yao Wu$^{a}$
\footnote{E-mail:wuxy2066@163.com}, Bai-Jun Zhang$^{a}$, Hai-Bo
Li$^{a}$  \\ Xiao-Jing Liu$^{a}$, Bing Liu$^{a}$, Jing-Wu Li$^{b}$
and Yi-Qing Guo$^{c}$}

\vspace{0.8cm}

\vskip 10pt \noindent{\footnotesize a. Institute of Physics, Jilin
Normal University, Siping 136000, China\\ \vskip 13pt
\footnotesize b. Institute of Physics, Xuzhou Normal University,
Siping 221116, China\\ \vskip 3pt \footnotesize c. Institute of
High Energy Physics, P. O. Box 918(3), Beijing 100049, China}

\end{center}

\date{}

\renewcommand{\thesection}{Sec. \Roman{section}} \topmargin 10pt
\renewcommand{\thesubsection}{ \arabic{subsection}} \topmargin 10pt
{\vskip 5mm
\begin {minipage}{140mm}
\centerline {\bf Abstract}
\vskip 8pt
\par
\indent \hspace{0.3in}We have analyzed the $D\rightarrow \pi K,
KK$ decay with the naive factorization (NF), QCD factorization
(QCDF)and QCD factorization including soft-gluon exchanges
(QCDF+SGE). In these decay channels, the soft-gluon effects are
firstly calculated with light cone QCD sum rules. Comparing the
three kind approaches, we can find the calculation results have
made much more improved QCD factorization (QCDF) than the naive
factorization (NF), and the calculation results have also made
improved QCD factorization including soft-gluon exchanges
(QCDF+SGE) than the QCD factorization (QCDF) in the
color-suppressed decay channels. In addition, we find the
soft-gluon effects are larger than the leading order
contributions, and the calculation results are close to the
experimental data for the color-suppressed decay channels. In
color-allowed decay channel $D^{0}\rightarrow \pi^{+} K^{-}$, the
soft-gluon effects are small and we should consider other power
terms, such as final state interaction and annihilation effects.

\end {minipage}

\vspace*{2cm} {\bf PACS number(s): 11.55.Hx, 11.15.Tk, 13.25.Ft}


\newpage
\section * {1. Introduction}
\hspace{0.3in}The study of heavy meson decays is important for
understanding the standard model (SM) and search for the sources
of $CP$ violation. However, The hadronic two-body weak decays of
$D$ meson involve nonperturbative strong interactions and spoil
the simplicity of the short distance behavior of weak
interactions. Therefore, a simplified approach in which the
amplitudes of these processes are given by a factorizable short
distance current-current effective Hamiltonian is not expected to
work well. Various approaches were employed to include long
distance effects. The most commonly and very frequently used
prescription, motivated by $\frac{1}{N_c}$ arguments \cite{s1}, is
to apply generalized factorization [2-3]. This phenomenological
treatment works reasonably well in color-allowed $D$ decays
\cite{s3}, but it is failing  in the color-suppressed
$D\rightarrow \pi \pi, \pi K$ and $D\rightarrow K \overline{K}$
decays \cite{s4}.

It is necessary that we study $D$ meson nonleptonic decays beyond
the factorization approach. A few years ago, M. Beneke et
al.\cite{s5} gave a NLO calculation the hadronic matrix element of
$B\rightarrow \pi \pi, \pi K$ in the heavy quark limit. They
pointed out that in the heavy quark limit the radiative
corrections at the order of $\alpha_{s}$ can be calculated with
perturbative QCD method. In $D\rightarrow \pi K$ decay, the
momentum transition square is $q^2=1.7GeV^2$, and the radiative
corrections of the hard-gluon exchanges can also be calculated
with perturbative QCD approach. So, the hadronic matrix elements
for $D\rightarrow \pi K$ can be expanded by the powers of
$\alpha_{s}$ and $\frac{\Lambda_{QCD}}{m_{c}}$ as follows:
\begin{equation}
\langle K \pi|O_{i}|D\rangle=\langle K|j_{1}|D\rangle \langle \pi
|j_{2}|0\rangle[1+\sum r_{n}
\alpha_{s}^n+O(\frac{\Lambda_{QCD}}{m_{c}})],
\end{equation}
where $O_{i}$ are some local four-quark operators in the weak
effective Hamiltonian and $j_{1,2}$ are bilinear quark currents.
In Eq. (1), the power correct term
$O(\frac{\Lambda_{QCD}}{m_{c}})$ includes soft-gluon effects,
final state interaction, which can not be calculated in QCD
factorization and perturbative QCD method. For the $B$ meson
two-body decay, the term is small, but it is large and can not be
neglected in the $D\rightarrow \pi K$ decay. A few years ago, A.
Khodjamirian \cite{s6} has presented a new method to calculate the
hadronic matrix elements of nonleptonic $B$ meson decays within
the framework of the light cone QCD sum rules, where the
nonfactorizable soft-gluon contributions can effectively be dealt
with. Obviously, this approach can be applied to $D\rightarrow \pi
K$ decay.

The QCD factorization method can be applied to
 $D\rightarrow\pi\pi, \pi K$ and $K\rho$ decay, but we should
 calculate the contribution of power term
 $O(\frac{\Lambda_{QCD}}{m_{c}})$. The power term includes the contributions
 of soft-gluon effect, final state interaction and annihilation
 effects, since the power term in $D\rightarrow\pi\pi, \pi K, K\rho$
 decay is larger than $B\rightarrow\pi\pi, \pi K, K\rho$ decay. We
 firstly considered $D\rightarrow\pi\pi$ decay in QCD
 factorization and light cone QCD sum rules method \cite{s7}. We found
 either the hard-gluon effect ($O(\alpha_{s})$ correction) or the
 soft-gluon effect is small, and only found the calculation result of
 $D^{0}\rightarrow\pi^{+}\pi^{-}$ decay channel approaches
the experiment data. It indicated that we should consider
 the contributions of final state interaction and annihilation
 effects in $D\rightarrow\pi\pi$ decay. In this paper, we apply the
 QCD factorization including light cone QCD sum rules method
 to study $D\rightarrow \pi K, KK$ decay and obtain new results. In $D^{0}\rightarrow
\pi^{0} \overline{K}^{0} $ decay, we find both hard gluon and soft
gluon contributions exceed the leading order largely, and the
 calculation result is accordance with the experiment data.
 In other decay channels, we
 should calculate all power terms, which include soft-gluon exchanges,
 final state interaction and annihilation
 effects, and then we can compare the
 calculation results with the experiment data. However, the final state interaction
 and annihilation effects haven't reliable method to calculate up to now. In our
work, we calculate the leading order and $\alpha_{s}$ corrections
in QCD factorization, and the soft-gluon effects in the light cone
QCD sum rules for the $D\rightarrow \pi K, K K$ decay. In
color-suppressed decay channels, we find the soft-gluon
contributions are larger than the leading order contributions, and
the calculation results are close to the experimental data for
these decay channels. In color-allowed decay channels, the
soft-gluon contributions are small and we should consider other
power terms, such as the final state interaction and annihilation
effects.

\section * {2. $D\rightarrow \pi K, KK$ in QCD Factorization}

\hspace{0.3in}The low energy effective Hamiltonian for
$D^{0}\rightarrow \pi^{0}\overline{K}^{0} $ can be expressed as
follows:
\begin{equation}
{\cal
H}_{eff}=\frac{G_{F}}{\sqrt{2}}V_{cs}^*V_{ud}[(C_{1}(\mu)O_{1}(\mu)
+C_{2}(\mu)O_{2}(\mu)],
\end{equation}
where $C_{i}(\mu)$ are Wilson coefficients which have been
evaluated to next-to-leading order. The four-quark operators
$O_{1,2}$
\begin{eqnarray}
&&O_{1}=(\bar{u}d)_{V-A}(\bar{s}c)_{V-A}, \nonumber  \\&&
O_{2}=(\bar{u}_{\alpha}d_{\beta})_{V-A}(\bar{s}_{\beta}c_{\alpha})_{V-A},
\end{eqnarray}
the Wilson coefficients evaluated at $\mu=m_c$ scale are \cite{s8}
\begin{equation}
C_{1}=1.274,    C_{2}=-0.529.
\end{equation}

In the following, we study $D \rightarrow \pi K, KK$ decay with
the QCD factorization approach. This method is similar to that for
$B\rightarrow \pi \pi, \pi K$ decay; see Ref. \cite{s5} for
detail. As in Ref. \cite{s5}, we obtain the QCD coefficients
$a_{i}$ at next-to-leading order (NLO) and $O(\alpha_s)$ hard
scattering corrections in naive dimension regularization (NDR)
scheme. The coefficient $a_{i}(\pi K) (i=1, 2)$ are split into two
terms: $a_{i}(\pi K)=a_{i,I}(\pi K)+a_{i,II}(\pi K)$. They are
given in Refs. [5, 14]. In $D\rightarrow \pi K, KK$ decay, the
flavor structure is different from $B$ decays. When we replace the
index $K$ and $\pi$ in the $B$ decays' coefficients $a_{i}(\pi K)
(i=1, 2)$, we can obtain the coefficients $a_{i}(\pi K) (i=1, 2)$
in $D$ decays. They are
\begin{eqnarray}
&&a_{1,I}=C_{1}+\frac{C_{2}}{N_{c}}(1+\frac{C_{F}\alpha_{s}}{4\pi}V_{\pi}),
a_{1,II}=\frac{C_{2}}{N_{c}}\frac{C_{F}\pi\alpha_{s}}{N_{c}}H_{\pi
K},
\end{eqnarray}
\begin{eqnarray}
&&a_{2,I}=C_{2}+\frac{C_{1}}{N_{c}}(1+\frac{C_{F}\alpha_{s}}{4\pi}V_{K}),
a_{2,II}=\frac{C_{1}}{N_{c}}\frac{C_{F}\pi\alpha_{s}}{N_{c}}H_{K
\pi}.
\end{eqnarray}
Here $N_{c}=3 (f=4)$ is the number of colors (flavors), and
$C_{F}=\frac{N_{c}^2-1}{2N_{c}}$ is the factor of color. The
functions in Eqs. (5) and (6) can be found in Ref. \cite{s5},
which are
\begin{eqnarray}
&&V_{K}=12\ln{\frac{m_{c}}{\mu}}-18+\int_{0}^{1}g(x)\phi_{K}(x)dx,
\nonumber \\&&
V_{\pi}=12\ln{\frac{m_{c}}{\mu}}-18+\int_{0}^{1}g(x)\phi_{\pi}(x)dx,
\nonumber \\&&
g(x)=3(\frac{1-2x}{1-x}\ln{x}-i\pi)+[2Li_{2}(x)-(\ln{x})^2+\frac{2\ln{x}}
{1-x}-(3+2i\pi)\ln{x}-(x\leftrightarrow1-x)], \nonumber \\&&
H_{\pi K}=\frac{f_{D}f_{K}}{m_{D}^2F^{D\rightarrow
K}(0)}\int_{0}^{1}\frac{\phi_{D}(\xi)}{\xi}d\xi\int_{0}^{1}\frac{dx}{\bar{x}}
\phi_{\pi}(x)\int_{0}^{1}\frac{dy}{\bar{y}}[\phi_{K}(y)+\frac{2\mu_{K}}{m_{c}}
\frac{\bar{x}}{x}],\nonumber \\&& H_{K \pi
}=\frac{f_{D}f_{\pi}}{m_{D}^2F^{D\rightarrow
\pi}(0)}\int_{0}^{1}\frac{\phi_{D}(\xi)}{\xi}d\xi\int_{0}^{1}\frac{dx}{\bar{x}}
\phi_{K}(x)\int_{0}^{1}\frac{dy}{\bar{y}}[\phi_{\pi}(y)+\frac{2\mu_{\pi}}{m_{c}}
\frac{\bar{x}}{x}],
\end{eqnarray}
where $Li_{2}(x)$ is the dilogarithm, $f_{K} (f_{D})$ is the $K$
($D$ meson) decay constant, $m_{D}$ is the $D$ meson mass,
 $F^{D\rightarrow K}(0)$ ($F^{D\rightarrow \pi}(0)$) is the $D\rightarrow K$ ($D\rightarrow \pi$) form factor at
zero momentum transfer, and $\xi$ is the light-cone momentum
fraction of the spectator in the $D$ meson. $H_{\pi K}$ and $H_{K
\pi}$ depend on the wave function $\phi_{D}$ through the integral
$\int_{0}^{1}d\xi\phi_{D}(\xi)/\xi\equiv m_{D}/\lambda_{D}=6.23$,
with $\lambda_{D}=(250 \pm 75)MeV$, $\mu_{K}=m_{K}^2/(m_d+m_s)$,
$\mu_{\pi}=m_{\pi}^2/(m_u+m_d)$, $m_u=3MeV$, $m_d=6MeV$,
$m_s=(150\pm 20)MeV$, $m_c=1.3GeV$, $m_\pi=0.139GeV$,
$m_K=0.494GeV$. We take $f_{\pi}=(132 \pm 0.26)MeV$, $f_{K}=(170
\pm 1.5)MeV$, $f_{D}=(200\pm 20)MeV$, $f_{D_{s}}=(280\pm 18)MeV$,
$F^{D\rightarrow \pi}(0)=(0.65\pm 0.10)$, $F^{D\rightarrow
K}(0)=(0.73\pm 0.07)$ \cite{s8}, $F^{D_{s}\rightarrow
K}(0)=(0.82\pm 0.15)$ \cite{s15}, $\alpha_{s}(m_{c})=0.353$,
$m_{D}=1.869GeV$, $m_{D_{s}}=1.968GeV$, and the asymptotic wave
functions $\phi_{K}=\phi_{\pi}=6x(1-x)$.

\newpage

\section * {3. $D\rightarrow \pi K, KK$ in the light-cone QCD sum rules}

\hspace{0.3in}In the following, we calculate the soft-gluon
contributions for $D\rightarrow \pi K, KK$ decay. Firstly, we
calculate the soft -gluon effects of $D^{0}\rightarrow
 \pi^{0}\overline{K}^{0}$ channel, and the calculation of other decay
channels are similar as the channel. To estimate the soft-gluon
corrections for $D^{0}\rightarrow  \pi^{0}\overline{K}^{0}$
channel, it is useful to rewrite down the effective Hamiltonian
with the help of the Fierz transformation. For example, applying
the Fierz transformation to the operator
$O_{2}=(\overline{u}\Gamma_{\mu}c)(\bar{s}\Gamma_{\mu}d)$, we have
the effective Hamiltonian relevant to the tree operators,
\begin{equation}
{\cal H}_{eff}=\frac{G_{F}}{\sqrt{2}}V_{cs}^*V_{ud}[(C_{1}(\mu)+
\frac{C_{2}(\mu)}
{3})O_{1}(\mu)+2C_{2}(\mu)\widetilde{O}_{1}(\mu)],
\end{equation}
where
\begin{equation}
\widetilde{O}_{1}=(\overline{u}\Gamma_{\mu}\frac{\lambda^a}{2}d)
(\overline{s}\Gamma^{\mu}\frac{\lambda^a}{2}c),
\end{equation}
In the above $\Gamma_{\mu} =\gamma_{\mu}(1-\gamma_{5})$,
$Tr(\lambda^a\lambda^b)=2\delta^{ab}$, and
\begin{equation}
O_{2}=\frac{1}{3}O_{1}+2\widetilde{O}_{1}.
\end{equation}

First of all, we calculate the nonfactorizable matrix elements
induced by the operator $\widetilde{O}_{1}$. As a starting object
for the derivation of LCSR we choose the following vacuum-pion
correlation function:
\begin{equation}
F_{\alpha}^{(\widetilde{O}_{1})}(p,q,k)=-\int d^4xe^{-i(p-q)x}\int
d^4ye^{i(p-k)}\langle0|
T\{j_{\alpha5}^{\overline{K}^{0}}(y)\widetilde{O}_{1}(0)j_{5}^{D}(x)\}|\pi^{0}(q)\rangle,
\end{equation}
where
$j_{\alpha5}^{(\overline{K}^{0})}=\overline{s}\gamma_{\alpha}\gamma_{5}d$
and $j_{5}^{(D)}=m_{c}\overline{c}i\gamma_{5}u$ are the quark
currents interpolating $\overline{K}^{0}$ and $D$ meson,
respectively. The decomposition of the correlation function (11)
in independent momenta is straightforward and contains four
invariant amplitudes:
\begin{equation}
F_{\alpha}^{(\widetilde{O}_{1})}=(p-k)_{\alpha}F^{(\widetilde{O}_{1})}+
q_{\alpha}\tilde{F_{1}}^{(\widetilde{O}_{1})}+
k_{\alpha}\tilde{F_{2}}^{(\widetilde{O}_{1})}+
\epsilon_{\alpha\beta\lambda\rho}q^{\beta}
p^{\lambda}k^{\rho}\tilde{F_{3}}^{(\widetilde{O}_{1})}.
\end{equation}
In what follows only the amplitude $F^{(\widetilde{O}_{1})}$ is
relevant. The correlation function is calculated in QCD by
expanding the T-product of three operators, two currents and
$\widetilde{O}_{1}$, near the light-cone $x^2 \sim y^2 \sim
(x-y)^2 \sim 0$. For this expansion to be valid, the kinematical
region should be chosen as:
\begin{equation}
q^2=p^2=k^2=0,     |(p-k)^2|\sim |(p-q)^2|\sim
|P^2|\gg{\Lambda_{QCD}}^2,
\end{equation}
where $P\equiv p-k-q$. The correlation function (11) can be
calculated employing the light-cone expansion of the quark
propagator $\cite{s6}$:
\begin{eqnarray}
S(x,0)&&= -i\langle0|T{q(x) \bar{q}(0)}|0\rangle \nonumber \\&&
=\frac{\Gamma(d/2)\hat{x}}{2\pi^2(-x^2)^{d/2}} \nonumber \\&&
+\frac{\Gamma(d/2-1)}{16\pi^2(-x^2)^{d/2-1}}
\int\limits_{0}^{1}dv((1-v)
\hat{x}\sigma_{\mu\nu}G^{\mu\nu}(vx)+v\sigma_{\mu\nu}G^{\mu\nu}(vx)\hat{x}),
\end{eqnarray}
where $G_{\mu\nu}=g_sG_{\mu\nu}^{a}(\lambda^{a}/2)$, which is the
gluonic field strength and the soft gluon effects are from the
term, $d$ is the space-time dimension. Following the standard
procedure for QCD sum rule calculation, we can obtain the hadronic
matrix element of the operator $\widetilde{O}_{1}$
\begin{eqnarray}
A^{(\widetilde{O}_{1})}(D^{0}\rightarrow \pi^{0} \overline{K}^{0})
=&&\langle
\overline{K}^{0}(-q)\pi^{0}(p) |\widetilde{O}_{1}|D^{0}(p-q) \rangle \nonumber \\
&& =\frac{-i}{\pi^2 f_{K}f_{D}{m_{D}}^2}\int_{0}^{s_{0}^{K}}dse^
{\frac{m_{K}^{2}-s}{M^2}}\int_{m_c^2}^{\bar{R}(s,m_c^2,m_{D}^2,s_0^{D})}
ds^{\prime} \nonumber \\ &&
e^{\frac{m_{D}^2-s^{\prime}}{{M^{\prime}}^2}}Im_{s^{\prime}}Im_{s}
F_{QCD}^{(\widetilde{O}_{1})}(s,s^{\prime},m_{D}^2),
\end{eqnarray}
where $s_{0}^{K}$ and $s_{0}^{D}$ are effective threshold
parameters for $K$ and $D$ meson.

A straightforward calculation shows that only the twist-3 wave
function $\varphi_{3\pi}(\alpha_{i})$ and the twist-4 ones
$\varphi_{\parallel}(\alpha_{i}),\varphi_{\perp}(\alpha_{i})$,
whose definitions can be found in Ref. \cite{s6}, contribute to
the invariant function $F^{(\widetilde{O}_{1})}$. The results are:
\begin{equation}
F^{(\widetilde{O}_{1})}_{QCD}=F_{tw3}^{(\widetilde{O}_{1})}+
F_{tw4}^{(\widetilde{O}_{1})},
\end{equation}
with
\begin{eqnarray}
F_{tw3}^{(\widetilde{O}_{1})}=&&\frac{m_c f_{3{\pi}}}{4
{\pi}^2}\int_{0}^{1} dv \int D\alpha_{i} \nonumber  \\ && \times
\frac{\varphi_{3 \pi}(\alpha_{i})}{(m_c^2-(p-q)^2(1-\alpha_{1}))
(-P^2v \alpha_{3}-(p-k)^2(1-v \alpha_{3}))} \nonumber  \\ &&
\times[(2-v)(q \cdot k)+2(1-v)q \cdot (p-k)](q \cdot (p-k)).
\end{eqnarray}
We get the same calculation results as Ref. \cite{s6} for Eq. (17)
when it be substituted for $m_c\rightarrow m_b$, but the twist-4
contribution hasn't been showed in Ref. \cite{s6}. Now, we give
the invariant amplitude from the twist-4 term.
\begin{eqnarray}
F_{tw4}^{(\widetilde{O}_{1})}=&&-\frac{m_c^2f_{\pi}}{4 {\pi}^2}
\int_{0}^{1}dv \int
D\alpha_{i}\widetilde{\varphi}_{\perp}(\alpha_{i})\frac
{1}{m_c^2-(p-q+q \alpha_{1})^2}\frac
{(4v-6)(p-k)q}{(p-k-qv\alpha_{3})^2} \nonumber \\ &&
+\frac{m_c^2f_{\pi}}{2 {\pi}^2}\int_{0}^{1}dv \int d\alpha_{1}
d\alpha_{3} \Phi_{1}(\alpha_{1},\alpha_{3})\frac {1}{[m_c^2-(p-q+q
\alpha_{1})^2]^2}\frac{(2pq-2vqk)(p-k)q} {(p-k-qv\alpha_{3})^2}
\nonumber \\&& -\frac{m_c^2f_{\pi}}{2 {\pi}^2}\int_{0}^{1}dv \int
d\alpha_{3} \Phi_{2} (\alpha_{3})\frac
{1}{[m_c^2-(p-q\alpha_{3})^2]^2}\frac{(2pq-2vqk)(p-k)q}
{(p-k-qv\alpha_{3})^2} \nonumber \\&& +\frac{m_c^2f_{\pi}}{2
{\pi}^2}\int_{0}^{1}dv2v^2\int d\alpha_{3} \Phi_{2}
(\alpha_{3})\frac
{1}{pq[m_c^2-(p-q\alpha_{3})^2]}\frac{[(p-k)q]^3}
{(p-k-qv\alpha_{3})^4} \nonumber \\&& -\frac{m_c^2f_{\pi}}{2
{\pi}^2}\int_{0}^{1}dv(2v-2)v\int d\alpha_{3} \Phi_{2}(\alpha_{3})
\frac {1}{m_c^2-(p-q\alpha_{3})^2}\frac{[(p-k)q]^2}
{(p-k-qv\alpha_{3})^4}.
\end{eqnarray}
In Eq. (17) and (18), we make use of the following nonlocal
operator matrix elements:
\begin{eqnarray}
&&\langle 0| \bar{u}(0)\sigma_{\mu\nu}\gamma_5G_{\alpha \beta}(v
y)u(x)|\pi^{0}(q)\rangle \nonumber \\ &&
=i\frac{f_{3\pi}}{\sqrt{2}}[(q_\alpha q_ \mu g_{\beta\nu} -q_\beta
q_\mu g_{\alpha \nu})-(q_\alpha q_\nu g_{\beta\mu}-q_\beta q_\nu
g_{\alpha\mu})] \nonumber \\ && \times \int
D\alpha_i\varphi_{3\pi}(\alpha _i,\mu) e^{- i q(x\alpha_1+yv
\alpha_3)},
\end{eqnarray}
\begin{eqnarray}
&&\langle 0| \bar{u}(0)i\gamma_\mu \widetilde{G}_{\alpha \beta}(v
y)u(x)|\pi^{0}(q)\rangle \nonumber \\ && =\frac{1}{\sqrt{2}}q_\mu
\frac{q_\alpha x_\beta-q_\beta x_\alpha}{qx} f_{\pi} \int
D\alpha_i\widetilde\varphi_\parallel(\alpha_i,\mu) e ^{-iq(x
(\alpha_1+y v \alpha_3)} \nonumber \\ && + \frac{1}{\sqrt{2}}
(g_{\mu\alpha}^{\perp}q_\beta-g_{\mu\beta}^{\perp}q_\alpha) f_\pi
\int D\alpha_i\widetilde\varphi_\perp(\alpha_i,\mu) e^{-iq(x
\alpha_1+y v \alpha_3)},
\end{eqnarray}
with $D\alpha_i=d\alpha_1 d\alpha_2 d\alpha_3
\delta(1-\alpha_1-\alpha_2 -\alpha_3)$. Finally, the LCSR for the
$D^{0}\rightarrow \pi^{0} \overline{K}^{0}$ matrix element of the
operator $\widetilde{O}_{1}$ from the soft-gluon exchange is
obtained by applying to the duality approximation and Borel
transformation in the $D$ channel. The result can be written as:
\begin{eqnarray}
A_{1}&=&A^{(\widetilde{O}_{1})}(D^{0}\rightarrow
\pi^{0}\overline{K}^{0}) \nonumber \\&=& im_{D}^2
(\frac{1}{4{\pi}^2 f_{K}}
\int_{0}^{s_{0}^{K}}dse^{\frac{m_K^2-s}{M^2}})
(\frac{m_c^2}{2f_{D} m_{D}^4}\int_{u_0^D}^{1} \frac{du}{u}
e^{\frac{m_{D}^2}{{M^{\prime}}^2}-\frac{m_c^2}{u{M^{\prime}}^2}}
\nonumber \\&&
\times[\frac{m_cf_{3\pi}}{u}\int_{0}^{u}\frac{dv}{v}\varphi_{3\pi}
(1-u,u-v,v)+f_{\pi}\int_{0}^{u}\frac{dv}{v}[3
\widetilde{\varphi}_{\perp} (1-u,u-v,v) \nonumber \\&&
-(\frac{m_c^2}{u{M^{\prime}}^2}-1)\frac{\Phi_{1}(1-u,v)}
{u}]+f_{\pi}(\frac{m_c^2}{u{M^{\prime}}^2}-2)\frac{\Phi_{2}(u)}{u^2}]),
\end{eqnarray}
where $u_0^D=m_c^2/s_0^D$, and the following definitions are
introduced:
\begin{eqnarray}
&&\frac{\partial {\Phi_{1}}(w,v)}{\partial w
}=\widetilde{\varphi}_{\perp}
(w,1-w-v,v)+\widetilde{\varphi}_{\parallel}(w,1-w-v,v), \nonumber
\\&& \frac{\partial {\Phi_{2}}(v)}{\partial v}=\Phi_{1}(1-v,v).
\end{eqnarray}
The asymptotic forms of the pion distribution amplitudes in Eqs.
(21)-(22) are given by \cite{s6}:
\begin{eqnarray}
&&\varphi_{3\pi}(\alpha_i)=\varphi_{3k}(\alpha_i)=360 \alpha_{1}
\alpha_{2} \alpha_{3}^2, \nonumber \\&&
\widetilde{\varphi}_{\perp}(\alpha_i)=10\delta^2\alpha_{3}^2
(1-\alpha_{3}), \nonumber \\&&
\widetilde{\varphi}_{\parallel}(\alpha_{i})=-40\delta^2\alpha_{1}
\alpha_{2}\alpha_{3}.
\end{eqnarray}
We find our calculation result (Eq. (21)) has a little difference
with Eq. (30) in Ref. [6]. In Eq. (21), the final term includes
function $(\frac{m_{c}^{2}}{uM'^{2}}-2)\frac{\Phi_{2}(u)}{u^{2}}$,
and this function is corresponding to
$(\frac{m_{c}^{2}}{uM'^{2}}-\frac{s}{M'^{2}}-1)\frac{\Phi_{2}(u)}{u^{2}}$
in Ref. [6], which includes variation $s$. We think the function
including variation $s$ should appear in the first integral
$\int_{0}^{s_{0}}dse^{\frac{-s}{M^2}}$, and it should not be in
the second integral $\int^{1}_{u_{0}}\cdots du$, so that the
matrix element $A$ is independent of variation $s$. Our
calculation result $A_{1}$ (Eq. (21)) is a constant. In Ref. [6],
the matrix element $A$ (Eq. (30)) is a function of variation $s$.
It should print error in Ref. [6].

We write the decay amplitudes of $D^{0}\rightarrow
\pi^{0}\overline{K}^{0}$, which include factorization and
non-factorization parts:
\begin{eqnarray}
M_{f+\alpha_{s}}(D^{0}\rightarrow \pi^{0}\overline{K}^{0})
=i\frac{G_{F}}{\sqrt{2}}V_{cs}^*V_{ud}f_{K} F^{D\rightarrow
\pi}(0) (m_{D}^2-m_{\pi}^2)a_{2},
\end{eqnarray}
\begin{eqnarray}
M_{nfg}(D^{0}\rightarrow \pi^{0}\overline{K}^{0})
=\sqrt{2}G_{F}V_{cs}^*V_{ud}C_{1}A_{1}.
\end{eqnarray}

The matrix elements of other decay channels, can be calculated as
similar as $D^{0}\rightarrow \pi^{0}\overline{K}^{0}$ channel, and
we can write down them directly as follows:
\begin{eqnarray}
A_{2}&=&A^{(\widetilde{O}_{1})}(D^{0}\rightarrow \pi^{+}K^{-})
\nonumber
\\&=& im_{D}^2 (\frac{1}{4{\pi}^2 f_{\pi}}
\int_{0}^{s_{0}^{\pi}}dse^{\frac{-s}{M^2}}) (\frac{m_c^2}{2f_{D}
m_{D}^4}\int_{u_0^D}^{1} \frac{du}{u}
e^{\frac{m_{D}^2}{{M^{\prime}}^2}-\frac{m_c^2}{u{M^{\prime}}^2}}
\nonumber \\&&
\times[\frac{m_cf_{3K}}{u}\int_{0}^{u}\frac{dv}{v}\varphi_{3K}
(1-u,u-v,v)+f_{K}\int_{0}^{u}\frac{dv}{v}[3
\widetilde{\varphi}_{\perp} (1-u,u-v,v) \nonumber \\&&
-(\frac{m_c^2}{u{M^{\prime}}^2}-1)\frac{\Phi_{1}(1-u,v)}
{u}]+f_{K}(\frac{m_c^2}{u{M^{\prime}}^2}-2)\frac{\Phi_{2}(u)}{u^2}]),
\end{eqnarray}
\begin{eqnarray}
A_{3}&=&A^{(\widetilde{O}_{2})}(D_{s}^{+}\rightarrow
K^{+}\overline{K}^{0}) \nonumber
\\&=& im_{D_{s}}^2 (\frac{1}{4{\pi}^2 f_{K}}
\int_{0}^{s_{0}^{K}}dse^{\frac{m_K^2-s}{M^2}})
(\frac{m_c^2}{2f_{D_{s}} m_{D_{s}}^4}\int_{u_0^D}^{1} \frac{du}{u}
e^{\frac{m_{D_{s}}^2}{{M^{\prime}}^2}-\frac{m_c^2}{u{M^{\prime}}^2}}
\nonumber \\&&
\times[\frac{m_cf_{3K}}{u}\int_{0}^{u}\frac{dv}{v}\varphi_{3K}
(1-u,u-v,v)+f_K\int_{0}^{u}\frac{dv}{v}[3
\widetilde{\varphi}_{\perp} (1-u,u-v,v) \nonumber \\&&
-(\frac{m_c^2}{u{M^{\prime}}^2}-1)\frac{\Phi_{1}(1-u,v)}
{u}]+f_K(\frac{m_c^2}{u{M^{\prime}}^2}-2)\frac{\Phi_{2}(u)}{u^2}]),
\end{eqnarray}
and
\begin{eqnarray}
A_{4}&=&A^{(\widetilde{O}_{2})}(D_{s}^{+}\rightarrow \pi K)
\nonumber
\\&=& im_{D_{s}}^2 (\frac{1}{4{\pi}^2 f_{\pi}}
\int_{0}^{s_{0}^{\pi}}dse^{\frac{-s}{M^2}})
(\frac{m_c^2}{2f_{D_{s}} m_{D_{s}}^4}\int_{u_0^D }^{1}
\frac{du}{u}
e^{\frac{m_{D_{s}}^2}{{M^{\prime}}^2}-\frac{m_c^2}{u{M^{\prime}}^2}}
\nonumber \\&&
\times[\frac{m_cf_{3K}}{u}\int_{0}^{u}\frac{dv}{v}\varphi_{3K}
(1-u,u-v,v)+f_K\int_{0}^{u}\frac{dv}{v}[3
\widetilde{\varphi}_{\perp} (1-u,u-v,v) \nonumber \\&&
-(\frac{m_c^2}{u{M^{\prime}}^2}-1)\frac{\Phi_{1}(1-u,v)}
{u}]+f_K(\frac{m_c^2}{u{M^{\prime}}^2}-2)\frac{\Phi_{2}(u)}{u^2}]),
\end{eqnarray}
where
$\tilde{O}_{2}=(\bar{u}\Gamma_{\mu}\frac{\lambda^{a}}{2}c)(\bar{s}
\Gamma_{\mu}\frac{\lambda^{a}}{2}d)$
and $s_{0}^{\pi}$ is effective threshold parameters for $\pi$
meson.

In the following, we give the decay amplitudes of other channels,
which include factorization and non-factorization parts:
\begin{eqnarray}
M_{f+\alpha_{s}}(D^{0}\rightarrow \pi^{+}K^{-})
=i\frac{G_{F}}{\sqrt{2}}V_{cs}^*V_{ud}f_{\pi} F^{D\rightarrow
K}(0) (m_{D}^2-m_{K}^2)a_{1},
\end{eqnarray}
\begin{eqnarray}
M_{nfg}(D^{0}\rightarrow \pi^{+}K^{-})
=\sqrt{2}G_{F}V_{cs}^*V_{ud}C_{2}A_{2}.
\end{eqnarray}
\begin{eqnarray}
M_{f+\alpha_{s}}(D^{+}\rightarrow \pi^{+}\overline{K}^{0})
&=&i\frac{G_{F}}{\sqrt{2}}V_{cs}^*V_{ud}[f_{\pi} F^{D\rightarrow
K}(0) (m_{D}^2-m_{K}^2)a_{1}\nonumber \\&&+f_{K} F^{D\rightarrow
\pi}(0) (m_{D}^2-m_{\pi}^2)a_{2}],
\end{eqnarray}
\begin{eqnarray}
M_{nfg}(D^{+}\rightarrow \pi^{+}\overline{K}^{0})
=\sqrt{2}G_{F}V_{cs}^*V_{ud}(C_{1}A_{1}+C_{2}A_{2}).
\end{eqnarray}
\begin{eqnarray}
M_{f+\alpha_{s}}(D_{s}^{+}\rightarrow K^{+}\overline{K}^{0})
=i\frac{G_{F}}{\sqrt{2}}V_{cs}^*V_{ud}f_{K} F^{D_{s}\rightarrow
K}(0) (m_{D}^2-m_{K}^2)a_{2},
\end{eqnarray}
\begin{eqnarray}
M_{nfg}(D_{s}^{+}\rightarrow K^{+}\overline{K}^{0})
=-\sqrt{2}G_{F}V_{cs}^*V_{ud}C_{1}A_{3}.
\end{eqnarray}
\begin{eqnarray}
M_{f+\alpha_{s}}(D_{s}^{+}\rightarrow \pi^{+}K^{0})
=i\frac{G_{F}}{\sqrt{2}}V_{cd}^*V_{ud}f_{\pi} F^{D_{s}\rightarrow
K}(0) (m_{D}^2-m_{K}^2)a_{1},
\end{eqnarray}
\begin{eqnarray}
M_{nfg}(D_{s}^{+}\rightarrow \pi^{+}K^{0})
=\sqrt{2}G_{F}V_{cd}^*V_{ud}C_{2}A_{4}.
\end{eqnarray}
where the amplitude $M_{f+\alpha_{s}}=M_{f}+M_{\alpha_{s}}$
represents the sum of the leading order factorization $M_{f}$ and
non-factorization $M_{\alpha_{s}}$ from the hard-gluon exchanges,
the amplitude $M_{nfg}$ is the non-factorization parts from
soft-gluon exchanges. The total amplitude $M$ is the sum of
$M_{f+\alpha_{s}}$ and $M_{nfg}$.

\section * {4. Numerical calculation}

\hspace{0.3in}In the numerical calculations we take
$s_0^{\pi}=0.7GeV^2$\cite{s6}, $s_0^{K}=1.2GeV^2$\cite{s9} and
$s_0^{D}=(6\pm 1)GeV^2$\cite{s10}. In $\mu_{c}=\sqrt {m_D^2-m_c^2}
\approx 1.3 GeV$, $f_{3\pi}(\mu_{c})=0.0035 GeV^2$,
${\delta}^2(\mu_c)=0.19 GeV^2$ are nonperturbative parameters in
light-cone wave functions \cite{s10}, and the CKM matrix are
$V^{*}_{ud}=V_{cs}=0.9734\div0.9749$ and $V_{cd}=0.227$
\cite{s11}. Having fixed the input parameters, one must find the
range of the values ${M^{\prime}}^2$ and $M^2$ for which the sum
rules (Eq. (21)) is reliable. At the interval ${M^{\prime}}^2=8-12
GeV^2$ and $M^2=6-15 GeV^2$, we find the value of $A_{1}$ (Eq.
(21)) is quite stable. The $D$ meson life time
$\tau{(D^{0})}=(4.12\pm0.027)\times 10^{-13}s$,
$\tau{(D^{+})}=(1.05\pm0.013)\times 10^{-12}s$,
$\tau{(D_{s}^{+})}=(4.9\pm0.09)\times 10^{-13}s$. In the $D$ rest
frame, the two body decay width is
\begin{equation}
\Gamma{(D\rightarrow P_{1}P_{2})}=\frac{1}{8\pi}{\vert
{M(D\rightarrow P_{1}P_{2})}\vert}^2\frac{\vert{P}\vert}{m_{D}^2},
\end{equation}
where $P_{1}$ and $P_{2}$ are two pseudoscalar meson ($\pi$ and
$K$), and the momentum of the $P_{1}$ meson is given by
\begin{equation}
\vert{P}\vert=\frac{[(m_{D}^2-(m_{P_{1}}+m_{P_{2}})^2)
(m_{D}^2-(m_{p_{1}}-m_{{P_{2}}})^2)]^{\frac{1}{2}}}{2m_{D}},
\end{equation}
The corresponding branching ratio is given by
\begin{equation}
Br(D\rightarrow P_{1}P_{2})=\frac{\Gamma{(D\rightarrow
P_{1}P_{2})}}{\Gamma_{total}}.
\end{equation}
where $\Gamma_{total}$ denotes the total decay width of $D$ meson.
The total decay width of one meson is related to its mean life
time $\tau$ by $\Gamma_{total}=\hbar/\tau$. With the above
parameters and formulae, we can get the branching ratios of
$D\rightarrow \pi K, KK$ decay
obtained in some approaches with that of the experiment. \\
Table 1: The branching ratios of $D\rightarrow \pi K, KK$ decay
obtained in some approaches together with experimental result.
\begin{center}
\begin{tabular}{|c|c|c|c|c|}
\hline
Decay channel & NF & QCDF & QCDF+SGE & Experiment
\\
\hline $D^{0}\rightarrow \pi^{0}\overline{K}^{0} $ & $2.4\times
10^{-3}$ & $(3.66\pm0.55)\times 10^{-2}$ &
$(2.20\pm0.11)\times 10^{-2}$ & $(2.28\pm0.22)\times 10^{-2}$ \\
\hline $D^{0}\rightarrow \pi^{+}K^{-} $ & $5.63\times 10^{-2}$ &
$(7.18\pm1.01)\times 10^{-2}$ &
$(6.15\pm0.40)\times 10^{-2}$ & $(3.80\pm0.09)\times 10^{-2}$ \\
\hline $D^{+}\rightarrow \pi^{+}\overline{K}^{0}$ & $6.55\times
10^{-2}$ & $(1.66\pm0.23)\times 10^{-2}$ &
$(2.70\pm0.17)\times 10^{-2}$ & $(2.77\pm0.18)\times 10^{-2}$ \\
\hline $D_{s}^{+}\rightarrow K^{+} \overline{K}^{0}$ & $2.15\times
10^{-3}$ & $(4.27\pm0.68)\times 10^{-2}$ &
$(3.86\pm0.26)\times 10^{-2}$ & $(3.6\pm1.1)\times 10^{-2}$ \\
\hline $D_{s}^{+}\rightarrow \pi^{+} K^{0}$ & $3.49\times 10^{-3}$
& $(5.64\pm0.97)\times 10^{-3}$ &
$(5.07\pm0.35)\times 10^{-3}$ & $<8\times 10^{-3}$ \\
\hline
\end{tabular}
\end{center}

The branching ratios of $D\rightarrow \pi K, KK$ decay channels
are presented in Table 1, where the second column is the result of
naive factorization (NF) and the total amplitudes
$M_{f+\alpha_{s}}$ in Eqs. (24), (29), (31), (33) and (35)
corresponding to different decay channels of $D\rightarrow \pi K,
KK$, $a_{1}$ and $a_{2}$ are calculated in the leading order, i.e.
the parameters $a_{1}=C_{1}+\frac{C_{2}}{3}$ and
$a_{2}=C_{2}+\frac{C_{1}}{3}$, the third column is the result of
QCD factorization (QCDF), the amplitudes are also calculated by
$M_{f+\alpha_{s}}$ in Eqs. (24), (29), (31), (33) and (35) but
$a_{1}$ and $a_{2}$ are calculated by QCD factorization approach
from Eqs. (5)-(7), which include the leading order and
$O(\alpha_{s})$ corrections, the fourth column is the result of
QCD factorization including soft-gluon exchanges (QCDF+SGE) which
are our results, the total amplitude is the sum of
$M_{f+\alpha_{s}}$ and $M_{ngf}$ in Eqs. (24)-(25) and (29)-(36 ),
and the final column is the experimental data \cite{s12}. From
Table 1, we can find that the prediction of naive factorization is
far from the experimental data and the QCD factorization method
have improved the calculation results. In our approach (QCD
factorization including soft-gluon effects), the calculation
results are close to the experimental data in $D^{0}\rightarrow
 \pi^{0}\overline{K}^{0}$, $D^{+}\rightarrow
\pi^{+}\overline{K}^{0}$ and $D^{+}_{s}\rightarrow
K^{+}\overline{K}^{0}$ decay channels. We find the soft-gluon
corrections are rather large, which are larger than the leading
order contributions in $D^{0}\rightarrow
 \pi^{0}\overline{K}^{0}$ and $D^{+}_{s}\rightarrow
K^{+}\overline{K}^{0}$ decay channels. For $D^{0}\rightarrow
\pi^{+}K^{-}$ decay, the results from the three approaches do not
agree with the experimental data, and we think the reason is that
we only calculate the soft-gluon contributions and do not consider
the final state interaction and the annihilation effects in the
power term $O(\Lambda_{QCD}/m_{c})$. The theoretical uncertainties
in the table 1 are estimated by the some parameters. We can give
the uncertainties from QCD factorization and soft-gluon effects,
respectively. In QCD factorization, the important uncertainties
are from the input parameters: form factors $F^{D\rightarrow K}$,
$F^{D_{s}\rightarrow K}$, $F^{D\rightarrow \pi}$, decay constants
$f_{D}$, $f_{D_{s}}$, $f_{K}$, $f_{\pi}$ and the parameters
$a_{1}$ and $a_{2}$. From Eqs. (5)-(7), we can find the
uncertainties of $a_{1}$ and $a_{2}$ are from: (1) The Wilson
coefficients: The coefficients $C_{1}$ and $C_{2}$ are uncertain,
and they are in the ranges of: $1.216 \sim 1.274$ and $-0.415 \sim
-0.53$ respectively. (2) The vertex corrections $V_{M}(M=\pi, K)$:
we need to input the light-cone distribution functions
$\phi_{M}(M=\pi, K)$. We take the asymptotic form of pion and kaon
light-cone distribution functions
$\phi_{\pi}(x)=\phi_{K}(x)=6x(1-x)$, but the accurate form should
be Gegenbauer polynomials. (3) The hard-scattering terms $H_{K
\pi}$ and $H_{\pi K}$: we need to calculate the moment
$\int_{0}^{1}\frac{d\xi}{\xi}\phi_{D}(\xi)=\frac{m_{D}}{\lambda_{D}}
$, the value of $\lambda_{D}$ at present is uncertain, a typical
range being $\lambda_{D}=(250\pm75)MeV$. The uncertainties from
QCD factorization less than $18\%$. In soft-gluon effects, the
uncertainties are from the input parameters: the decay constants
$f_{D}$, $f_{D_{s}}$, $f_{K}$, $f_{\pi}$, the Wilson coefficients
$C_{1}$ and $C_{2}$, and the Borel parameters ${M^{\prime}}^2$ and
$M^2$. The uncertainties from soft-gluon effects less than $7\%$,
and the total uncertainties less than $25\%$. In $D\rightarrow \pi
K, KK$, we find the QCD factorization (QCDF) and QCD factorization
including soft-gluon exchanges (QCDF+SGE) have improvement on the
calculation results. So the QCD factorization method can be
applied to $D\rightarrow \pi K, KK$ decay, but the power term
$O(\Lambda_{QCD}/m_{c})$ should be included. In order to get
accurate calculation results in $D\rightarrow \pi K, KK$ decay,
the soft-gluon, the final state interaction and the annihilation
effects should be calculated, and the calculation results can be
improved. Now, the final state interaction and the annihilation
effects have many models, but there is not a reliable method. So,
it is necessary to study the $D\rightarrow \pi K, KK$ decay
further.

\section * {5. Summary}
\hspace{0.3in}In $D\rightarrow \pi K, KK$ decay, We find the
prediction of naive factorization (NF) is far from the
experimental data, the QCD factorization (QCDF) and QCD
factorization including soft-gluon exchanges (QCDF+SGE) have
improvement on the calculation results. In color-suppressed decay
channels, such as $D^{0}\rightarrow
 \pi^{0}\overline{K}^{0}$ and $D^{+}_{s}\rightarrow K^{+}\overline{K}^{0}$ decay, the
soft-gluon corrections are rather large, which are larger than the
leading order contributions. In color-allowed decay channel
$D^{+}\rightarrow \pi^{+}\overline{K}^{0}$, the soft-gluon
corrections are rather large also, but the soft-gluon corrections
are small in the color-allowed decay channel $D^{0}\rightarrow
\pi^{+} K^{-}$. From Refs.[6, 13], we can find the soft-gluon
contributions are different in different $B\rightarrow \pi \pi$
decay channels. For example, in $B^{0}\rightarrow \pi^{+} \pi^{-}$
decay, the soft-gluon effects are smaller than the hard-gluon
contributions because of the small Wilson coefficient $C_{2}$ in
the decay amplitude. However, the soft gluon effects and the hard
gluon are on the same order in $B^{0}\rightarrow \pi^{0} \pi^{0}$
decay since the Wilson coefficient $C_{1}$ is large in this decay
amplitude. In $D\rightarrow \pi K, KK$ decay, we can obtain the
similar results. The QCD factorization method can be applied to
$D\rightarrow \pi K, KK$ decay, but the power term
$O(\Lambda_{QCD}/m_{c})$ should be included. In order to get
accurate calculation results in $D\rightarrow \pi K, KK$ decay,
the soft-gluon, the final state interaction and the annihilation
effects should be calculated, and then the calculation results can
be improved.

\newpage
\end{document}